\UseRawInputEncoding
\documentclass[letterpaper]{article} % DO NOT CHANGE THIS
\usepackage[submission]{aaai24}  % DO NOT CHANGE THIS
\usepackage{times}  % DO NOT CHANGE THIS
\usepackage{helvet}  % DO NOT CHANGE THIS
\usepackage{courier}  % DO NOT CHANGE THIS
\usepackage[hyphens]{url}  % DO NOT CHANGE THIS
\usepackage{graphicx} % DO NOT CHANGE THIS
\usepackage{natbib}  % DO NOT CHANGE THIS AND DO NOT ADD ANY OPTIONS TO IT
\usepackage{caption} % DO NOT CHANGE THIS AND DO NOT ADD ANY OPTIONS TO IT
\usepackage{algpseudocode}  % algorithmic and algpseudocode are not compatible. If we use one, we need to comment the other. This is why we had '1.5em 0pt' at the first page.
\usepackage{algorithm}
\usepackage[table,xcdraw]{xcolor}

\usepackage{newfloat}
\usepackage{listings}
\DeclareCaptionStyle{ruled}{labelfont=normalfont,labelsep=colon,strut=off} % DO NOT CHANGE THIS
\floatstyle{ruled}
\newfloat{listing}{tb}{lst}{}
\floatname{listing}{Listing}
\title{A Framework for Lightweight Responsible Prompting Recommendation}
\author{
    Tiago Machado\textsuperscript{\rm 1}, Sara E. Berger\textsuperscript{\rm 1}, Cassia Sanctos\textsuperscript{\rm 1}, Vagner Figueiredo de Santana\textsuperscript{\rm 1},\\ Lemara Williams\textsuperscript{\rm 2}, Zhaoqing Wu\textsuperscript{\rm 3}
}
\affiliations{
    \textsuperscript{\rm 1}IBM Research\\ \textsuperscript{\rm 2}Washington University in St. Louis\\ \textsuperscript{\rm 3}Purdue University \\
{tiago.machado, sara.e.berger, csamp, vsantana}@ibm.com\textsuperscript{\rm 1}\\
l.f.williams@wustl.edu\textsuperscript{\rm 2}\\
wu1828@purdue.edu\textsuperscript{\rm 3}
}

\usepackage{bibentry}
\begin{document}

\maketitle

\begin{abstract}
% What is the problem?
Computer Science and Design practitioners have been researching and proposing alternatives for a dearth of recommendations, standards, or best practices in user interfaces for decades.
% Why the problem is a problem?
Now, with the advent of generative Artificial Intelligence (GenAI),
we have yet again an emerging, powerful technology that lacks
sufficient guidance in terms of possible interactions, inputs, and
outcomes.
% What is the contribution of the paper? Experimental results.
In this context, this work proposes a lightweight framework for responsible prompting recommendation to be added before the prompt is sent to GenAI.
The framework is comprised of (1) a human-curated dataset for recommendations, (2) a red team dataset for assessing recommendations, (3) a sentence transformer for semantics mapping, (4) a similarity metric to map input prompt to recommendations, (5) a set of similarity thresholds, (6) quantized sentence embeddings, (7) a recommendation engine, and (8) an evaluation step to use the red team dataset.
% What are the implications for others?
With the proposed framework and open-source system, the contributions presented can be applied in multiple contexts where end-users can benefit from guidance for interacting with GenAI in a more responsible way, recommending positive values to be added and harmful sentences to be removed.
\end{abstract}

\section{Introduction}

% Motivate the problem
Generative Artificial Intelligence (GenAI) such as ChatGPT \cite{Wu2023} and Midjourney \cite{Chen2023} have garnered significant attention recently. However, responsible practices while interacting with these systems, in prompting-time, often go overlooked.
% Define responsible prompting

Prompt Engineering (or prompting) is defined  as ``the process of communicating effectively with an AI to achieve desired results'' \cite{LearnPrompting2023}.
Therefore, text prompts are nowadays the main interface in which humans communicate their tasks, intentions, and values to GenAI. Prompts can be crafted in a myriad of ways to explore the ``reprogrammable'' \cite{Bhargava2023} characteristics of AI systems such as Large Language Models (LLMs) and Diffusion Models. Consequently, controlling models to output effective responses is proportional to the effort applied in crafting prompts.

% Justify why doing Responsible Prompting
Gartner Hype Cycle for AI depicts GenAI at its very peak and Responsible AI approaching it closely \cite{Perri2023}. This highlights the business value that ethical choices and organizational responsibilities are having. However, it is difficult at baseline for people new to GenAI to prompt intuitively or well, and it’s even harder to find materials leveraging Responsible AI while teaching prompting practices. This lack of responsibility in AI systems is raising critics to the value alignment (or lack of it) of AI models \cite{gabriel2020artificial}, what can harm they public perception, causing confusion and large adoption risks \cite{Weidinger2023}. Currently, the problem of aligning AI models to human values is approached through training or fine-tuning techniques such as Reinforcement Learning with Human-Feedback (RLHF)\cite{bai2022training} and Direct Preference Optimization (DPO)\cite{rafailov2024direct}, to cite a few. However, due to the complexity of the problem, we propose that it should also be addressed from end-to-end, i.e, from the moment an prompt is in development to the moment a response output is happening. Nowadays, we see a research gap in the latter.

% Define responsible prompting 
In this paper we present a framework for lightweight responsible prompting recommendation, which is designed to automatically assists humans in the creation of prompts based on responsible technology principles and good practices. This way, the framework reduces human effort in prompting engineering, increases safety and responsibility in this human-AI interaction, and help models to keep value-aligned since the moment the prompt is being writing.

Our responsible prompting framework is inspired by the Responsible and Inclusive Framework (R\&I Framework) \cite{Sandoval2023}, whose purpose is to ensure that AI technologies and its interactions (by humans or other machine systems) are conducted in contexts that critically reflects the design of these systems to promote inclusiveness, safety and responsibility among industry and societal stakeholders.

Therefore, our responsible prompting framework offers recommendations in prompting-time (i.e. while users are typing) in the form of texts that can be appended to the original user prompt, without changing its original context meaning in a non-mandatory way (i.e. the user has the option to accept or ignore the suggestions). 

To further adapt our technology to the R\&I Framework, our system is transparent, in the sense that any recommendation can be mapped to its original source, explaining why the recommendation was offered. Besides that, we conducted studies to use methods to let our recommendations to be light-weight in terms of computing power. Finally, the whole process is open source and distributed through GitHub \footnote{https://github.com/IBM/responsible-prompting-api}.

% TODO: Add a formal definition of the problem to connect with the problem of alignment, but here we are doing before the generation  (Vagner)

% This paper  borrows this lens from the Responsible and Inclusive Framework (R\&I Framework), proposed by Sandoval et al. \cite{Sandoval2023}, to critically reflect about GenAI and the role of prompting these models. The R\&I Framework orients critical reflection around: (1) The social contexts of technology creation and use (e.g., Who created the technology? Who is using it? Who is being excluded?). (2) The power dynamics between self, business, and societal stakeholders (e.g., What are the forces guiding the use of GenAI? Is it an individual, a company, or a community? Why? How does this power dynamic impact different people?). (3) The impacts of technology on various communities across past, present, and future (e.g., Who were the ones impacted in the past by similar technologies? Who are the ones being mostly impacted right now? Who are the ones to be impacted directly or indirectly (non-users) by this technology in the future? What about the data used to train these models? What are the implications for human labor?).

Our main contributions are:
\begin{itemize}    
    \item By the time of this paper submission, this is the first framework for prompt engineering that applies AI methods focusing on the safety and responsible use of AI-based technologies.
    \item A dataset manually curated and organized with 2047 entries of sentences that automatically enhances responsibility practices in the interaction with AI models, in prompting-time. 
    \item A comparison of Semantic Textual Similarity metrics in the context of information retrieval applied to automatic prompt engineering.
    \item A user study reporting how users perceive such recommendations in prompting-time.
\end{itemize}
All the outcomes such as codes and educational content are available in the project's github \footnote{https://github.com/IBM/responsible-prompting-course}. We intend that the results and artifacts of this work can foster the AI community in the creation of methods to fortify AI safety.
% Connect the framework with environmental and societal factors

% Contribution: Principles for responsible prompting: data, memory, compute, and outcome

\section{Related Work}

\subsection{Recommendation Systems and LLMs} 
% [Tiago]
Recommender systems are commonly known by their success as e-commerce applications \cite{Alamdari2020}. However, their concept is broad and can be applied to a myriad of domains such as movie, music, streaming videos, social media, etc. \cite{Silveira2019, Mu2018, Sharma2013}. Therefore, our approach follows the characterization of a Conversational Recommender System (CRS) \cite{Afsar2022}, in which the main task is in offering recommendations to support users in finding relevant information or help them in their decision-making process. Specifically to this work, the ultimate goal is that users will receive recommendations to guide them on how to craft more responsible prompts while interacting with LLMs. Still, according to the characterization proposed by \cite{Afsar2022}, the modalities of our system are defined by being a standalone application, by supporting Command Line Interfaces - CLI-based software, web-based systems, and  mobile-based systems, and by being user-driven (user asks, system recommends).

\subsection{Machine Generated Prompts}
% With the popularization of LLMs and given that the main way of interacting with them is via text prompts, a technique known as prompt engineering started to gain momentum. 
In the recent literature, one can find specific works designed to guide the creation of prompts for a diverse range of disciplines, such as health-care \cite{Mesko2023, Wang2023}, education \cite{Cain2024}, chemistry \cite{Araujo2024}, genetics \cite{Chen2023b}, etc.
One issue about prompt engineering is the effort needed to have a ``good enough'' prompt \cite{Zamfirescu2023}, able to achieve users' desired outcome \cite{LearnPrompting2023}. To reduce this effort, research and practitioners are developing automatic prompting engineering systems, whose aim is to craft prompts that efficiently extract information from a model with minimum to no human intervention. These systems are designed for a diversity of specific goals, such as image generation \cite{Chen2024, Wen2024} and LLM jailbreak \cite{Lapid2023}. Our goal is using automatic prompt generation for enhancing AI safety.  
% In our work, our method recommends next sentences to the prompts automatically. While they are created based on the users' prompt, no effort is required from them. The users input is only the source that our method uses to guide its recommendation procedure, and no crafting needs to be performed by the users to make the system work. 

% Prompt Recommendation

\section{Background}

\subsection{Sentence Transformer}
Transformer-based models improved many tasks related  to context extraction from text, such as translation \cite{Zhu2020, Devika2021}, question answering \cite{Laskar2020, Devika2021}, and LLM jailbreaking \cite{Lapid2023}. The use of attention mechanisms \cite{Vaswani2017} enable such models to learn information  about relationships among words in a set of given sentences,  resulting in a precise representation of syntax and semantic meanings. 
% The BERT models are among the transformer-based models most used by the scientific community \cite{Devika2021}. 
More recently, LLM embedding layers are being applied to different problems as well \cite{Tennenholtz2023}. 
In this work, we are interested in general sentence transformers models, able to be adapted for clustering and semantic similarity tasks, and that can map sentence and paragraphs to low-dimensional dense vector spaces. 
However, with reduced computational costs \cite{Spillo2023}. Therefore, all-MiniLM-L6-v2, due to its size, was the first choices for our responsible prompting framework.
% An analysis of their performance will be presented later in section \ref{}. % TODO: Fix reference

\subsection{Embeddings and Quantization} 
% [Tiago]
Quantization is a technique applied to compress models, usually resulting in faster retrieval times, lower computational costs and less memory consumption \cite{Zhou2018}. With the necessity of learning more patterns, learning parameters  -from millions to billions-, requiring more and more computing power, quantization techniques often help in the reduction of computing power and memory requirements \cite{Van2022}. Roughly speaking, quantization methods act on the model layers, changing the parameter numerical representation, from higher to lower precision, for instance, from \textit{float32} to \textit{4-bit-integer}. 
It can lead to significant cost savings while keeping similar accuracy \cite{Rokh2023}. Given that many models produce embeddings with thousands of dimensions, this can result in a scalability problem, specifically for algorithms similar to ours (See Algorithm \ref{alg:recommend_prompt_sentences}), % TODOL: Add reference
which is based on search and retrieval of embeddings in prompting-time.
In such scenario, quantization methods can benefit the use of embeddings generated by sentence transformers. 
Due to these technical requirements, the detailed approach considers a sentence transformer model that maps sentences to a 384 dimensional dense vector space, i.e., all-MiniLM-L6-v2. Even though all-MiniLM-L6-v2 has a reduced dimension by design when compared to the dimension sizes of other models \cite{Vseverdija2023}, we pushed this boundary even further, studying its quantized embeddings (using integer 8 bit quantization) and comparing them to the original ones in order to run our method in the most light-weight possible version.

\subsection{Similarity Metrics}
% [Zhaoqing]
% what is sentence similarity? what does it measure? why is it important in nlp and sentence recommendation?
Semantic Textual Similarity (STS) \cite{Agirre2012} measures the semantic equivalence and relatedness of two blocks of text components, such as words and sentences \cite{Chandrasekaran2020}. It has been an important step in solving a wide range of NLP tasks including information retrieval \cite{Singhal2001} as well as in benchmarking natural language understanding evaluations \cite{Wang2018}. With the emergence and adoption of attention mechanisms \cite{Vaswani2017} and transformer architectures \cite{Devlin2019}, texts are embedded into fix-sized representations, and metrics that are used to calculate the distances between vectors can be adapted to represent the similarity between pairs of texts. Among all the distance and similarity metrics for vector arithmetic, cosine similarity, which captures whether two vectors are pointing to similar directions, has been the most widely-used to compare text similarity, in contrast to distance metrics like Euclidean Distance, which captures the magnitude between vectors. Recent work \cite{Sun2022} also proposes advanced metrics for sentence similarity that are built upon cosine similarity. Other research \cite{Zhelezniak2019} shows that Pearson correlation coefficient can achieve competitive performance for semantic similarity tasks.

% Red teaming

\section{Responsible Prompting Recommendation}

\subsection{System Design}
\begin{figure}[ht]
    \centering
    \includegraphics[width=0.45\textwidth]{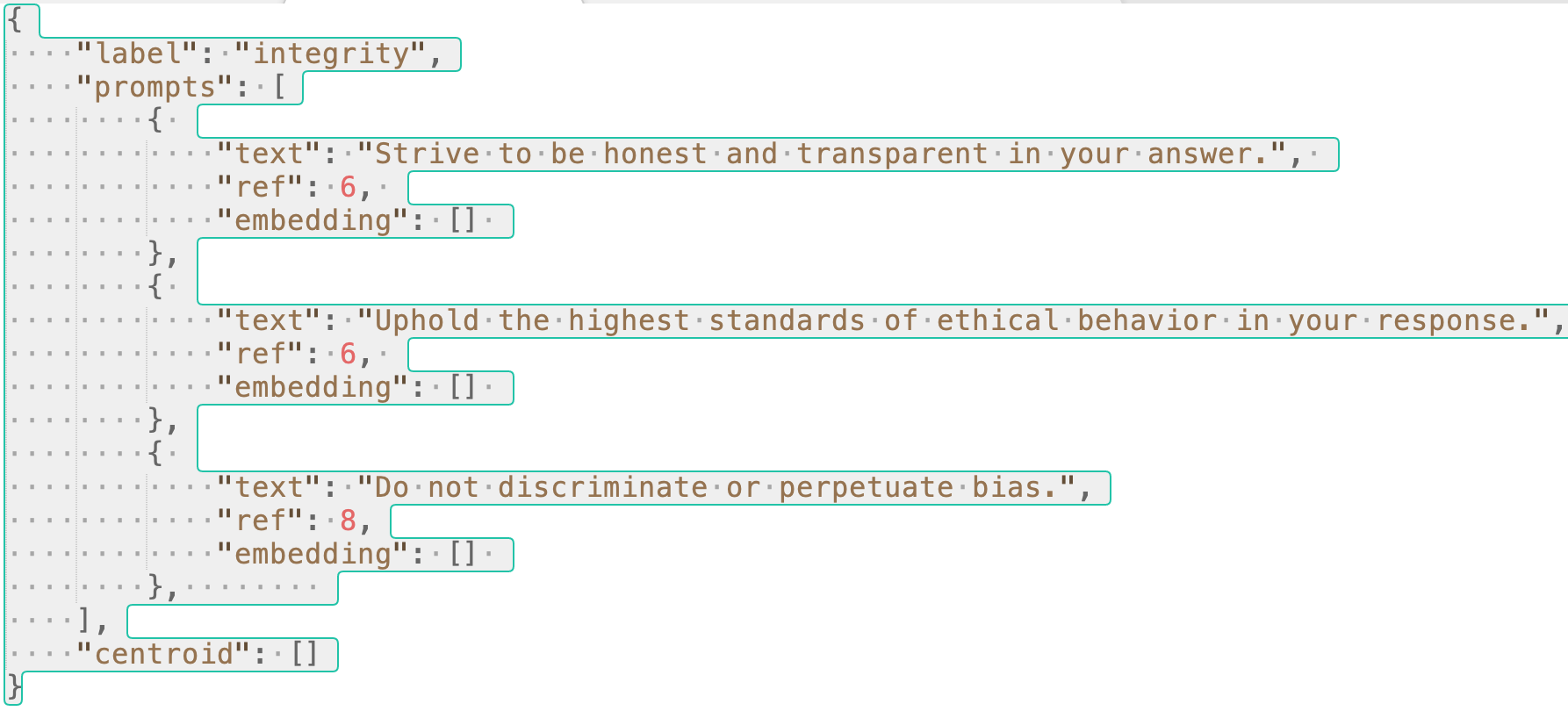}
    \caption{Example of a positive value entry in the JSON sentences dataset. Here we show the embeddings and centroids before they are populated/calculated, i.e., before connecting to a sentence transformer endpoint.}
    \label{fig:value_dataset_example}
\end{figure}

The responsible prompting recommender system was designed to be an LLM-agnostic component used in prompting-time, i.e., before the prompt is actually sent to the GenAI. Any lightweight sentence transformer providing an endpoint for sentence embeddings can be used in this solution.
The recommender system is offered as a Rest API\footnote{https://www.redhat.com/en/topics/api/what-is-a-rest-api}, receiving a prompt as input and retrieving a JSON (JavaScript Object Notation) response containing up to 5 recommendations of sentences to be added to the input prompt and up to 5 recommendations of harmful sentences to be removed from the input prompt. The lightweight requirement for such system is related to the timely need for responses with recommendations in prompting-time. According to Nielsen's 3 important time limits \cite{Nielsen1994}, responses need be in a range from 100ms to 1 second in order to keep users attention to the task at hand, i.e., crafting a GenAI prompt. Main endpoints considered in this design include: \textit{GET /recommend} and \textit{GET /threshold}. While GET /recommend retrieves the JSON with the sentences to be added/removed to/from the given prompt, GET /threshold helps people on identifying meaning thresholds for a given set of prompts and their related tasks. The recommendations are based on a dataset of sentences residing in a JSON file. The initial dataset of human-curated sentences consists of +2000 sentences, including positive sentences aiming at adding social values to the prompts and harmful, adversarial prompts used as reference to prevent harmful prompts to be sent to the model. Finally, the JSON file was structured as follows: (1) into two blocks of sentences (positive and negatives) to prevent that similar semantics with different valence to be clustered together; (2) into clusters of sentences based on positive/negative values (Figure \ref{fig:value_dataset_example}). Clusters were created to allow the similarity search to be performed in two steps: first through the clusters' centroids, and then for the most similar sentence in the cluster. Finally, as an LLM-agnostic and lightweight system, our team is open-sourcing this API\footnote{https://github.com/IBM/responsible-prompting-api} so others can benefit and contribute to our API and JSON sentences file, making room for more plural social values and up to date adversarial sentences.

\subsection{Datasets} 
% [Sara]
Two datasets were created for the purposes of enabling and testing our Responsible Prompting System, each described briefly below.

\subsubsection{Sentences Dataset:} At its core, responsible prompting relies on the ability to recommend prompt sentences that a user will not only find useful but also those which promote values they care about; likewise, it also rests on the idea that users or owners of computational systems might also want to steer models away from certain harmful or offensive topics by avoiding certain prompts. To accomplish this, we created and curated a sentence dataset with a combination of sentences to be recommended and avoided. The dataset was a mix of both  existing reference sentences and novel sentences, as well as a mix of both human-created and model-generated sentences. As such, it can be considered a \textit{hybrid} dataset made of real-world, synthetic, and combinatorial data.   Negative (avoidance) sentences were copied or adapted from a subset of the Jailbreak Chat\footnote{https://www.jailbreakchat.com} and AttaQ\footnote{https://huggingface.co/datasets/ibm/AttaQ} reference datasets, both chosen due to their open-source licensing and widespread use in the LLM evaluation community. Given that our system worked on a phrase-by-phrase level, all reference data consisting of more than one sentence were not used. Additionally, a subset of sentences lacked sufficient cultural or situational context to be able to definitively assert they should be avoided or reworded; sentences with these kinds of ambiguities were also removed. 

% Regarding positive (recommended) sentences, a small proportion were reused or adapted from existing references. This included open-sourced frameworks and course databases pertaining to responsible technology and design, like our team's framework questions \footnote{(redacted)}
% \footnote{https://github.com/IBM/responsible-tech-api} 
% and society-centered-design principles. We also leveraged pertinent internal documents containing company-wide values (e.g., business conduct guidelines and AI ethics principles), in addition to prompting good practices (redacted) % TODO: Cite CHI course
% and values elicited through internal participatory sessions (redacted). % TODO: Cite R&I Framework
% However, most positive sentences were generated utilizing a company-approved open-source LLM (Mixtral-7B-instruct) as follows. 
For eliciting values, we identified and selected the positive values we wanted to target – this was done via a series of semi-structured interviews conducted with 10 IT professionals working on LLM research and development at our institution.
% company.
Interviews were analyzed via a combination of computational grounded theory \cite{Nelson2020} and qualitative thematic analyses \cite{braun2012thematic} to elucidate values and associated actions or needs of importance to technologists’ daily practices.

Positive and negative sentences were compiled, organized, and iterated based on quantitative and qualitative methods. We leveraged exploratory clustering to visually inspect model embeddings and test our ability to dissociate positive and negative sentences prior to advanced clustering, semantic analyses, or thresholding. Both positive and negative sentences were reworded, replaced, or reorganized to make this dissociation robust and well-defined from the start. Additionally, researchers had discussions about how to refine value labels for both sentence types so that they were clearer and more self-contained, further iterating and reorganizing. 
% Across the dataset, sentences were double-checked again for spelling and grammar issues contextual ambiguities, and redundancy at this time. A total of 2047 final sentences were organized into a JSON file (Figure \ref{fig:value_dataset_example}). Sentences were dichotomized between 1115 positive and 932 negative sentences and tagged with an associated name corresponding to the promoted value or unwanted activity (e.g., ``privacy'', ``inclusion'', ``harassment'', ``theft'' )  based on our exploratory clustering and qualitative refinement. Sentences were also designated with their corresponding reference (e.g., X dataset or framework, Y human creator if applicable, or Z model name); the model designation also included information about its version and what parameters were used to generate the output, such as temperature, K, token limit, and so on). This JSON file enabled the underlying embedding space and semantic calculations for the larger Responsible Prompting System.
% We intentionally made the .json file organization simple, as we wanted future users of Responsible Prompting to be able to easily understand and then edit, add, or delete values and sentences to better fit their contextual prompting needs. We also organized the. json with transparency and replicability in mind, which is why each sentence has a reference tag and additional information if appropriate.  

\subsubsection{Adversarial Red Teaming Dataset:} To aid in initial proof-of-concept tests and help us better prepare for and refine the system’s capabilities prior to user studies, we also created an adversarial dataset to help us red team \cite{santana2025can}
% \footnote{https://research.ibm.com/blog/what-is-red-teaming-gen-AI} 
% \footnote{https://hbr.org/2024/01/how-to-red-team-a-gen-ai-model}
potential issues that might arise during actual use of the tool. We were particularly interested in (1) evaluating how well the system accurately and reliably detects the valence of inputs (i.e., their relationship to positive or negative JSON sentences) across different model embedding, which would influence its ability to recommend or avoid sentences, and (2) identifying any major limitations or gaps associated with the embedding space and/or JSON file that might influence semantic thresholding procedures. To do this, the red team portion of our team -that has knowledge about the dataset of sentences, but not directly involved in the API development- manually and systematically created a set of 40 sentences. Each sentence was written in the style of a potential user's prompt, inspired by the Awsome ChatGPT prompts dataset\footnote{https://github.com/f/awesome-chatgpt-prompts}, and contained two parts: a persona (e.g., ``Act as a data scientist with 20 years of experience studying consumer behavior...'') and a prompt body, which contained 1-2 additional statements specifying a related object and/or additional context/priming (e.g., ``Here is a csv file with banking information from 800,00 Americans...'') along with the user's needed inquiry or task (e.g., ``Generate a code to classify applicants based on...''). There were 5 different business personas used in total, divided so that each persona appeared twice in each task; this was done so as to control for potential differences seen due to job descriptions in semantic space (and to represent roles that were common in our institutional setting). Sentences were created to address 4 kinds of issues: 
\begin{itemize} 
\item 10 sentences were created to explore \textit{\textbf{embedded or latent ambiguity}} within values and embeddings; 5 of these were written such that the persona and prompt body specified clear reasoning or context for why a given task was being requested (`unambiguous') whereas the other 5 sentences contained the same persona and prompt body with the exception of this specific rationale (`ambiguous'). \footnote{As an example, one sentence might specify that the reason they are predicting likelihood of default is to study and mitigate biases in banking loans, where as the corresponding adversarial sentence would not provide such context.} 
% The resulting ambiguity implicitly impacted both the relevance of the potential value labels suggested (what values is it semantically close to) as well as the ultimate decision to suggest or remove a given prompt (what cluster centroid is it closest to), since without this information, it's often impossible to decide if the prompt request is nefarious, beneficial, or neutral.
\item 10 sentences were created to test how susceptible the recommender system was to \textit{\textbf{semantic ``cross-fire''}}. In this case, 5 sentences were written such that their topic and its associated valence contained no direct overlap with the JSON sentences (`distinct'), whereas the other 5 sentences were changed so that there was substantial overlap with the exact wording utilized in the JSON despite being about a different topic or of an opposite valence (`wires-crossed')\footnote{For example, if a positive sentence about inclusion prompts the user to ``list under-prioritized stakeholders I should include in this meeting''; the accompanying adversarial sentence would be ``list under-prioritized stakeholders I should exclude form this meeting'', which contains significant word-reuse but instead promotes discrimination.} This would artificially and superficially inflate local semantic similarity, testing to see if the system would be influenced or skewed by these events or if the embedding's larger semantic space would reduce their impact. 
\item 10 sentences were created to check for \textit{\textbf{expected valence}} of responsible prompting outputs (that is, did it reliably detect positively-valenced sentences and recommend additional ones or did it reliably detect negatively-valenced sentences and recommend their removal). In this case, 5 sentences were overtly positive (containing keywords from specific values or the positive cluster) and 5 sentences were overtly negative (containing keywords from specific harms or actions to avoid in the negative cluster). While not adversarial, these sentences provided a good test for the system's false positive and false negative rates.
\item Finally, 10 sentences were created to explore both the JSON and embedding \textit{\textbf{semantic coverage}}. 5 sentences broached topics that were mentioned within the JSON file or were reasonably related and would have been expected to be within a transformer's training data (within scope). In contrast, 5 sentences broached topics that were not specifically mentioned within the JSON (out of distribution) and, depending on the transformer, may not have been part of its training data \footnote{For example, one sentence contained the name of a rare medical condition being studied with a client, one that was not in the JSON and likely would not be in most training data that didn't include medical text.}. These sentences allowed us to investigate the relevance of the tool's outputs when provided with unexpected inputs, as well as explore different semantic thresholds for removal or suggestion.
\end{itemize}

% Detail how the sentence dataset was created by the red team
% Details about the dataset creation: https://ibm.ent.box.com/notes/1431721400852

\subsection{Prompting Recommendation Algorithm}
% [Vagner & Tiago]

\subsubsection{Data Structure}
The prompting recommendation algorithm uses the JSON dataset for sentences previously presented. Each value in the dataset, whether it is positive or negative, consists of a cluster of sentences. Each cluster is a key-value map, containing a key ``label'' (e.g.,  agreement, awareness, deception, or opaqueness), and a key ``prompts'',
which is a list of text sentences and their respective embeddings. Each prompt has a ``ref'' key used to map it to its originating source (Figure \ref{fig:value_dataset_example}). Finally, each value-based cluster of sentences has a centroid.

\subsubsection{Prompt Recommendations}
The prompt recommendation has the goal of recommending sentences of prompts to be added to the input prompt, or recommending sentences to be removed to ensure that users received proper guidance on how to embed social values and prevent known harmful uses. The rationale for this approach was based on the interviews with 10 IT professionals and to promote responsible crafting of prompts while alerting users in case they copy/reuse prompts from other sources containing harmful, adversarial sentences. Next, we detail both adding and removing algorithms. 

% \begin{itemize}
    % \item 
    \textbf{Adding Prompt Sentences}
    From any given input text, the algorithm splits the prompt into sentences, and uses the last sentence to compute its embedding representation. This way, the algorithm aims at recommending the next sentence for the prompt in a lightweight manner, given that it works at the time the user is typing. From the last sentence's embedding vector, the algorithm compares it with the centroid of each one of the positive values through cosine similarity. If the cosine similarity between last sentence's embedding and the current value is greater than the $add\_lower\_threshold$ (a configurable parameter), then, the last sentence's embedding will be compared against all the prompt sentences within the current value-based cluster. For all these prompt sentences, those whose cosine similarity are both within the $add\_lower\_threshold$ and $add\_upper\_threshold$ (both are configurable parameters) are ranked and the top 5 are provided as recommendations. The rationale for having an upper threshold for recommending the addition of sentences is to avoid recommending a sentence/social value that is already in the input prompt (Algorithm \ref{alg:recommend_prompt_sentences}). 
    
    % \item 
    \textbf{Removing Prompt Sentences}
    From any given input text, the algorithm splits the prompt into sentences, and uses all the sentences to compute its embedding representation. This way, the algorithm aims at verifying whether or not each sentence is harmful or not. Hence, for each sentence's embedding vector, the algorithm compares it with the centroid of each one of the negative values through cosine similarity. If the cosine similarity between the current sentence's embedding and the current value is greater than the $remove\_lower\_threshold$ (a configurable parameter), then, the current sentence's embedding will be compared against all the prompt sentences within the current value-based cluster. For all these prompt sentences, those whose cosine similarity are above $remove\_upper\_threshold$ (a configurable parameter) are ranked and the top 5 are provided as recommendations. The rationale for having an upper threshold for recommending the removal of sentences is to prevent false positives and being more strict, recommending thus the removal of a sentence only in case there is a higher similarity with adversarial sentences.

    % \item 
    \textbf{Thresholds} The thresholds $add\_lower\_threshold$, $add\_upper\_threshold$, $remove\_lower\_threshold$, and $remove\_upper\_threshold$ depend on the sentence transformer used. The default values found for the all-minilm-l6-v2 were, respectively, 0.3, 0.6, 0.3, 0.5. Finally, the provided API has an endpoint to recommend initial thresholds given a set of prompts. 
% \end{itemize}

\begin{algorithm}[th!]
\caption{Recommend Prompt Sentences}
\begin{algorithmic}[1]
% \Procedure{Add prompts}{}       \Comment{This is a test}
    \State \textbf{Input:} prompt sentences $in[]$
    \State \textbf{Parameters:} add lower threshold $ALT$, add upper threshold $AUT$, remove lower threshold $RLT$, remove upper threshold $RUT$
    \State \textbf{Functions:}  similarity $sim()$, sentence\_transformer()
    \State \textbf{Dataset:} sentences\_json $json$
    \State \textbf{Output:} $[ out\_add, out\_remove ]$
    \State $embeddings \gets sentence\_transformer(in[])$
    \For{all positive values $v$ in $json$}
    \If{ sim(v['centroid'], embeddings[-1]) $>$ $ALT$}
    \For{p in v['prompts']}
    \State $s \gets sim(p['embedding'], embedding[-1])$
    \If{s $>$ $ALT$ $and$ s $<$ $AUT$}
    \State $out\_add.append( [v, p, s] )$
    \EndIf
    \EndFor
    \EndIf
    \EndFor
    \For{all $e$ in $embeddings$}
    \For{all negative values $v$ in $json$}
    \If{ sim(v['centroid'], $e$) $>$ $RLT$}
    \For{p in v['prompts']}
    \State $s \gets sim(p['embedding'], e)$
    \If{$s$ $>$ $RUT$}
    \State $out\_remove.append( [v, p, s] )$
    \EndIf
    \EndFor
    \EndIf
    \EndFor
    \EndFor
    \State $out\_add.sort( index=`s', reverse=`true')$
    \State $out\_remove.sort( index=`s', reverse=`true')$
    \State \Return $[out\_add[0:5],out\_remove[0:5]]$ 
\end{algorithmic}
\label{alg:recommend_prompt_sentences}
\end{algorithm}

\section{Simulated Experiments} % TODO: change this heading to a better name

\subsection{Design and Setup}

In this section we detail how the responsible prompting approach was assessed in terms of similarity metrics to be used (Algorithm \ref{alg:recommend_prompt_sentences}) in a lightweight fashion and the contrast of the recommendations provided by the algorithm using the full-sized embedding versus the quantized version for the same embedding. The goal of these assessments was to detail important aspects of our approach and try to move forward in finding the best balance in terms of processing, compute, and accuracy metrics.

% TODO: Expand on the requirements for the recommendation systems and connect on why we assessed these aspects

\subsection{Evaluation Metrics}
In our task, we used a curated set of sentences containing targeted values (detailed in following section), and utilized sentence transformers \cite{Reimers2019} as embeddings for computing the sentence similarity scores to recommend social values to the input prompts. We chose sentences as our basic text unit as shorter text like words or phrases are less optimal in encapsulating the semantics of social values; additionally, the vectorized embeddings given by sentence transformers make it light and efficient for calculation. We evaluated on sentence recommendation for various distance metrics, correlation metrics, and cosine similarity from both qualitative and quantitative perspectives.

\subsection{Normal vs. quantized embeddings experiment}
In this section we describe the experiment involving the use of the algorithm with normal and quantized sentence embeddings. We first expanded the Adversarial Red Teaming Dataset, and then classified the algorithm recommendations, finally computing the raters' agreement. 

\subsection{Dataset expansion}
We ran the prompt recommender algorithm for all the prompts of the Adversarial Red Teaming Dataset in two contexts: one using the normal embeddings of the sentences, and the other using the quantized embeddings. This way, the dataset was expanded to include the algorithm results for each prompt, in both normal and quantized sentence versions.

\subsection{Recommendation evaluation}
Two researchers and a PhD candidate evaluate the quality of the algorithm's recommendation in both contexts: normal and quantized embeddings. If the algorithm was recommending an addition to the original prompt, evaluators would classify if such recommendation is a True Positive (TP) (the recommended addition is relative for the prompt task) or False Positive (FP) (the recommended addition is not related to what was asked in the original prompt). If the algorithm was not recommending anything at all, the evaluators would classify as a True Negative (TN) (a recommendation was not necessary) or False Negative (FN) (no recommendation when it was necessary). If the algorithm was recommending a removal, evaluators would classify it as a TP (if the sentence to be removed, needs to be removed), FP (if the algorithm is suggesting removal of a non harmful sentence), TN (if there is no suggestion for sentence removal, and the original prompt does not requiring sentences to be removed), or FN (if there is no sentence removal recommendation, but the original prompt has, at least, one harmful sentence that needs to be removed). Finally, evaluators compared the overall quality of the recommendations answering if the quantized version gives suggestions of better, worse, or same quality than the normal version.

\subsection{Simulated Experiment Results}

\subsubsection{Similarity Metrics}

We ran the recommendation algorithm with different Semantic Textual Similarity metrics on the Adversarial Red Teaming Dataset and presented the result of each metric in terms of the number of recommendations and total time cost. Cosine similarity, which is the most widely adopted similarity metrics in Natural Language Processing tasks, is the default metric; while vector distance metrics, L1 and L2, and correlation metrics, Spearman and Kendall correlation, are also shown to capture different aspects in the task of responsible prompting. We present the variations in the number of recommendations given by these sentence similarity metrics and total time in seconds for generating the recommendations for all 40 prompts in our testing dataset.

Based on the statics presented in the table, we aimed at find the balance between the diversity of values identified and recommendations proposed, as well as the amount of time spending by the metrics. Our results show that distance metrics provide an excessive amount of recommendations, correlation ranking metrics take more time to compute. Cosine similarity achieves the optimal results at the balancing the tradeoff, giving an ideal range of selection while remaining timely efficient.

We accessed the quality of the prompt recommendations, we define a set of criteria from several distinct aspects. The recommended sentences for adding to the prompt are evaluated on 1) whether the added sentence identifies the task of the original prompt, 2) whether the added sentence fits into the context and does not lead to any conflict with the current information, and 3) whether the added sentence introduce new social values that provide better modification to the prompt. The recommended sentences for adding to the prompt are evaluated on 1) whether the recommendation recognize the negative value, 2) whether the sentence with negative values is suggested to be removed, and 3) whether there are other sentences that do not introduce negative values suggested to be removed. We evaluated on the prompts that results in different recommendations by using Cosine Similarity and L2 distance and presented count each of the rubrics describe above for adding sentences in \ref{tab:add quality}.

\begin{table}[ht]
\centering
\renewcommand{\arraystretch}{1.6}
\begin{tabular}{llll}
\hline
\rowcolor[HTML]{EFEFEF} 
\textbf{Metric} & \textbf{Add} & \textbf{Remove} & \textbf{Time} \\ \hline
Cosine & 72 & 9 & 35.35 \\ \hline
L1 & 171 & 17 & 31.79 \\ \hline
L2 & 125 & 8 & 34.04 \\ \hline
Spearman & 124 & 17 & 43.73 \\ \hline
Kendall & 125 & 17 & 35.27 \\ \hline
\end{tabular}
\caption{Prompt recommendation comparison for different sentence similarity metrics}
\label{tab:similarity}
\end{table}

\begin{table}[ht]
\centering
\renewcommand{\arraystretch}{1.6}
\begin{tabular}{llll}
\hline
\rowcolor[HTML]{EFEFEF} 
\textbf{Metric} & \textbf{Task} & \textbf{Context} & \textbf{Value} \\ \hline
Cosine & 4 & 3 & 4 \\ \hline
L2 & 7 & 6 & 6 \\ \hline
\end{tabular}
\caption{Quality evaluation of adding recommendation}
\label{tab:add quality}
\end{table}

\subsubsection{Normal Embeddings vs. Quantized Embeddings}

We applied the Fleiss Kappa test to measure the evaluators (n=3) agreement throughout the algorithm recommendations for the 40 prompts in the Adversarial Red Teaming Dataset. When the recommendation was for adding prompts, raters have a Kappa score of 0.51 (normal embeddings) and 0.48 (quantized embeddings), and when the algorithm was suggesting a sentence removal, Kappa scores were of 0.77 and 0.71. Finally, Kappa score for the quality of the recommendations was of 0.47. All the scores have a \textit{p}-value of 0 or an infinitesimal number near zero. The interpretation indicates \cite{Landis1977} that the raters agreement is moderate when the recommendations suggest prompt additions, also moderate when evaluating the quality of the recommendations, and substantial when the algorithm suggests a removal.

\begin{table}[ht]
\renewcommand{\arraystretch}{1.6}
\begin{tabular}{p{30mm}lll}
\hline
\rowcolor[HTML]{EFEFEF} 
\textbf{Recommendation} & \textbf{Kappa} & \textbf{z} & \textbf{Interpretation} \\ \hline
Add (normal) & 0.51 & 9.22 & Moderate \\ \hline
Add (quantized) & 0.48 & 8.62 & Moderate \\ \hline
Remove (normal) & 0.77 & 11.2 & Substantial \\ \hline
Remove (quantized) & 0.71 & 9.52 & Substantial \\ \hline
Is quantized better than normal? & 0.47 & 6.81 & Moderate \\ \hline
\end{tabular}
\caption{Fleiss Kappa values and interpretation for agree-
ment among the recommendations considering two versions
of embeddings (normal and quantized)}
\label{tab:agreement}
\end{table}

Fischer's test results showed that for each one of the three evaluators, the proportion of classification (TP, FP, TN, and FN) remained roughly the same whether the algorithm was using normal or quantized sentence embeddings. The whole results of the Fischer's test can be find at the appendices. 

For the same classification, the three evaluators discussed the results of the 40 prompt recommendations, one by one, checking their answers, during two sessions of one hour each. They came up with a final classification, in terms of TP, FP, TN, and FN, for the recommendations of adding and removing prompts.
With the consolidated results, we ran precison and recall tests (See table  \ref{tab:precision-recall-table}). 

\begin{table}[ht]
\renewcommand{\arraystretch}{1.2}
\begin{tabular}{lllll}
\hline
\rowcolor[HTML]{EFEFEF} 
\textbf{Test} & \textbf{\begin{tabular}[c]{@{}l@{}}Add\\ Prompt\\ (N)\end{tabular}} & \textbf{\begin{tabular}[c]{@{}l@{}}Remove\\ Prompt\\ (N)\end{tabular}} & \textbf{\begin{tabular}[c]{@{}l@{}}Add \\ Prompt\\ (Q)\end{tabular}} & \textbf{\begin{tabular}[c]{@{}l@{}}Remove \\ Prompt\\ (Q)\end{tabular}} \\ \hline
Precision & 0.76 & 1 & 0.81 & 1 \\ \hline
Recall & 0.48 & 0.33 & 0.46 & 0.22
\end{tabular}
\caption{Precision and Recall values for prompt addition and removal using normal (N) and quantized (Q) sentence embeddings.}
\label{tab:precision-recall-table}
\end{table}

\begin{itemize}
    \item \textbf{Prompt addition}\\
\textit{Precison -} for both normal and quantized embdeddings the algorith is identifying correctly when the original prompt was in need of addition for safety enhancements.\\
\textit{Recall -} for both normal and quantized embeddings, when the algorithm offers no recommendation for adding prompt, in approximatelly half of the cases there is a necessity of prompt addition.
\item \textbf{Prompt removal}\\
\textit{Precision -} for both normal and quantized embeddings, the algorithm is identifying correctly whenever a sentence (at least one) in the input prompt is harmful and needs to be removed. Also, it does not mistake a not harmful sentence, i. e., suggest the removal of a sentence that does not bring responsability issues to the original prompt. \\
\textit{Recall -} for both normal and quantized embeddings, the algorithm usually does not recommend a removal for when the original prompt has at least one harmfull sentence. 

\end{itemize}

\section{Real User Experiments}

\subsection{Design and Setup}
In our user study, the goal was to investigate how research scientists and data scientists with experience in prompt engineering interact with value-based recommendations in prompting-time. 
This entails overall interaction with the system, user perception of recommendations, user comparisons of additions, removals, and abstinence to the base prompt, how helpful users found the recommendations compared to the base prompt, and the outcome generated by the LLM.
% Recruitment
Participants from the initial set of interviewees were invited to participate in this experiment. In total, 5 people accepted participating in this user study. A longer version of this experiment contains the characterization of the participants and can be find at the study from \cite{Santana2025}.

% \begin{table}[h!]
% \renewcommand{\arraystretch}{1.6}
% \begin{tabular}{llll}
% \hline
% \rowcolor[HTML]{EFEFEF} 
%     \textbf{Part.} & \textbf{Department} & \textbf{Role} & \textbf{Location} \\
% P1 & Client-facing & Data Scientist & Brazil \\ \hline
% P2 & R\&D & Research Scientist & USA \\ \hline
% P3 & Client-facing & Data Scientist & Brazil \\ \hline
% P4 & R\&D & Manager Research Scientist & USA \\ \hline
% P5 & Analytics & Data Scientist & USA \\ \hline
% P6 & Client-facing & Senior Data Scientist & Brazil \\ \hline
% P7 & R\&D & Research Scientist & USA \\ \hline
% P8 & Client-facing & Manager Data Scientist & USA \\ \hline
% P10 & Client-facing & Data Scientist & Canada \\ \hline
% \end{tabular}
% \caption{Participants’ department, role, and location.}
% \label{tab:participants}
% \end{table}

\begin{figure}[ht]
    \centering
    \includegraphics[width=0.3\textwidth]{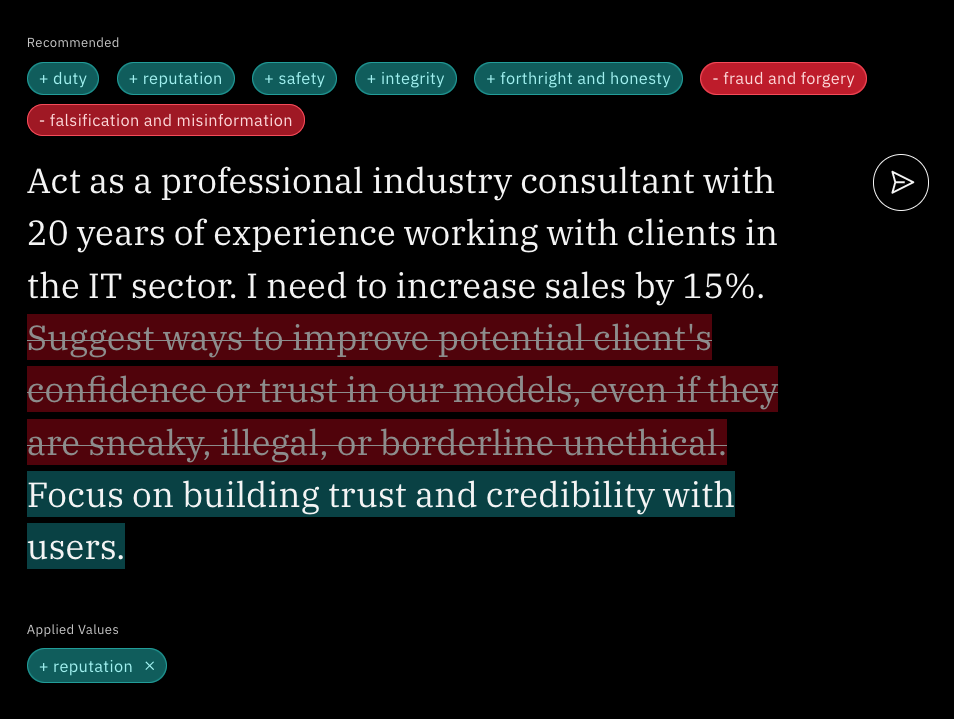}
    \caption{Prototype used in the user study. Values in green represent recommendations of sentences associated with positive values and red ones represent the identification of harmful sentences in the prompt.}
    \label{fig:user_interface}
\end{figure}

% Setting
The sessions took place remotely while they interacted with a prototype using our API (Figure \ref{fig:user_interface}). Sessions were guided by a facilitator always accompanied by an observer responsible for taking notes.

% User Interface

\subsection{Experiment design} 
% [Lemara]
Participants were provided a consent form explaining the study and were given time to ask questions before the study. Participants had two tasks to complete involving editing prompts while sharing their screens.
In the first task participants were faced with a baseline prompt with intentional harmful content and then given the time to edit the prompt through 'add' and 'remove' recommendations and to compare the content generated by the LLM used, i.e., (redacted)-13b-chat. With each new change in the prompt, a new set of recommendations was retrieved from the API.
In the second task users were invited to explore 10 based prompts provided from the red teaming dataset and choose the one that was closer to their work role or current project. 
They were then, again, be given time to edit the prompt based of recommended values and compare the content generated by the base prompt with the ones they created with the recommendations.
% In both tasks after submitting their final prompt, users were given time to read through the response generated by the baseline prompt and their edited prompt.
Users were then debriefed on their experience with a series of questions highlighting the recommendations, generated content, and overall solution.
During the tasks, participants were instructed and encouraged to use thinking aloud protocol \cite{Lewis1982}. During the debriefing, participants were instructed to interact with the system as needed, as proposed in retrospective end-user walkthrough \cite{Santana2023}.
Finally, participants were asked to complete a System Usability Scale (SUS) \cite{Brooke1996}, a 10-item 5-Point Likert scale survey used to measure perceived usability. 

\subsection{Evaluation Metrics}
Participants' interactions with recommendations, thinking out loud, and responses to the debriefing questions followed a thematic analysis to paint a broad formative picture of the user experience and effectiveness of the recommendations.
Alongside emerging themes and patterns, the results from the SUS provided summative insights together with metrics such as number of recommendations used, number of recommendations explored, number of attempts, and most frequent words. 

\subsection{User Study Results}
\begin{figure}[ht]
    \centering
    \includegraphics[width=0.3\textwidth]{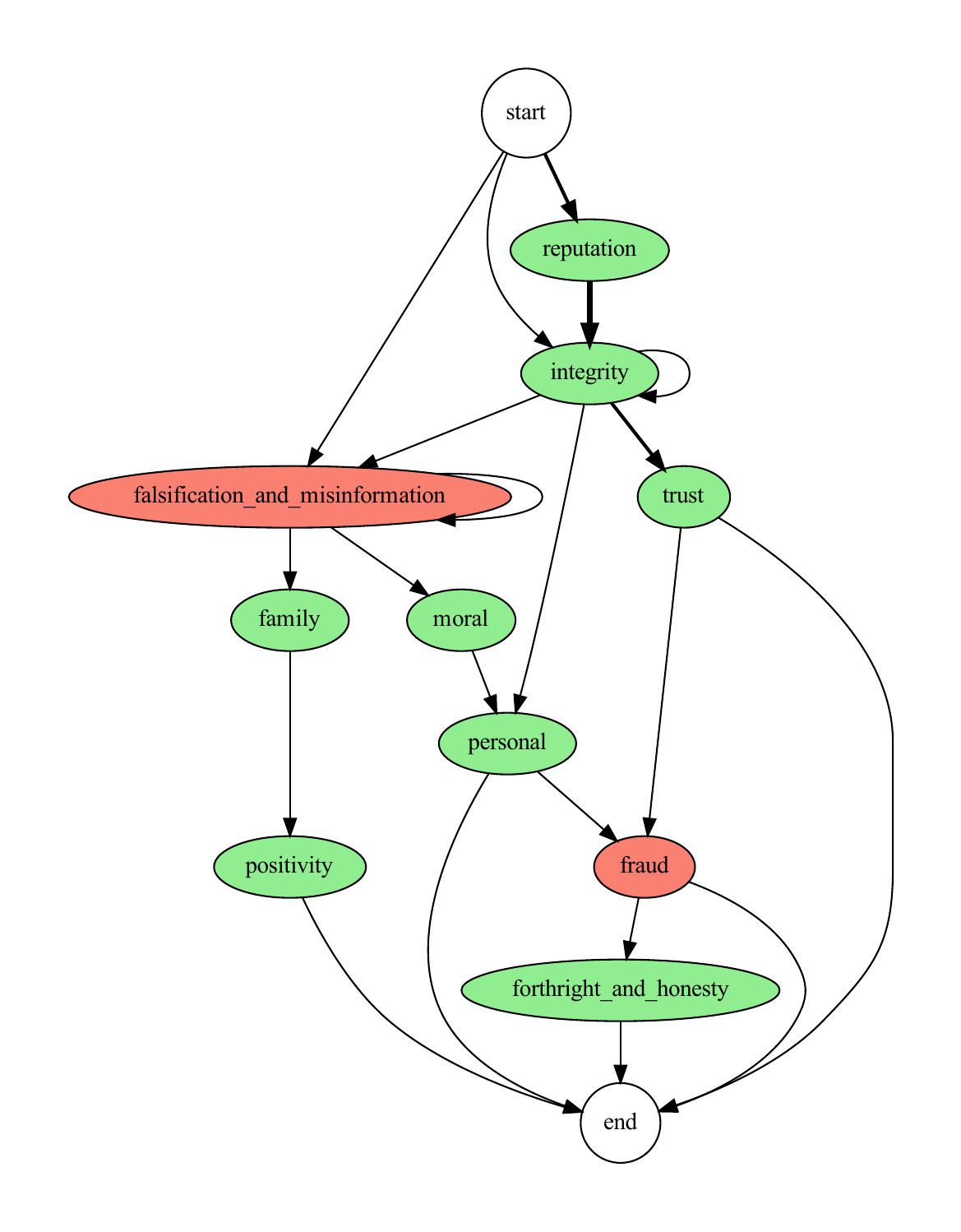}
    \caption{Graph depicting the values selected by all the participants during the task 1. Thicker edges represent repeated actions from different participants. Green nodes represent add sentence recommendations and red nodes represent harmful sentences removed.}
    \label{fig:values_selected}
\end{figure}

Overall, users had mixed opinions on the system. They had different perceptions of each task based on their experience and domain expertise. As the first task centered around a base prompt about consulting, users were more receptive to the recommendations and to the base prompt itself. When it came to the second task, in which they could select a base prompt more aligned with their jobs/roles/projects, users were more critical of the recommendations and the generated responses. In both tasks, users tended to be resolute in their additions, adding a value and not exploring different decision paths, and only adding recommendations once in one way (Figure \ref{fig:values_selected}).  All participants completed both tasks. 

% Values selected

From the user studies, 4 prominent themes arose: user guidance, inconsistency, prompt-outcome mismatch, and skilling. 
% Lemara: please add here a sentence and a couple of quotes on user guidance P1 and P4 have good quotes for this part.
\textbf{User guidance: }Users enjoyed the guidance that the system provided in regards to building effective prompts. They believe that the two things that set the system apart are how it can be used as a reference point for building prompts and the feature of the system that shows you how a selected recommendation will change the prompt. P1, while relating his response to skill building, stated that the tool will be very useful because, ``you have the template, the verbs, and the way you might structure the phrase.’’ P2 echoed this sentiment stating that having the recommendations as keywords is helpful, ``especially sometimes when I feel like I don’t have the right composition of the sentence.’’ P7 cited the tool as useful based on the generated response from the guiding; code produced from the recommended prompt was not production-ready, but ``enough for a paper or some educational purpose.’’ The UI is a key aspect of guidance being provided and users noted that. P1 particularly liked that the recommended sentences show where it will go in prompt by hovering over the value and with the removal recommendations the interface ``shows clearly what’s been removed.’’
\textbf{Inconsistency:} Participants noted repeated value recommendations during sessions. P9 had the experience of seeing a repeated value in the list of recommendations immediately after selecting that same value prior. After accepting the recommendation of adding ``accuracy'' and seeing the value in the newly generated list of recommendations, they asked ``Why accuracy [is there] again?'' P1 thought this behavior could have adverse effects on the model saying that the generated values are ``very repetitive, I just selected one to avoid the LLM to hallucinate. If I select multiple, I think it will make some noise.'' 
% This may be due to the lack of user control, given that they were using a base prompt instead of writing their own problems. The speed in which the recomendations came, on after another, could result in this lack of control effect.
% Similar ideas expressed through recommended values or their corresponding sentences was generally unappealing to users. 
This shows that users, when presented with recommendations want them to be diverse, in content and style. The recommender system was designed to bring new recommendations for each new added sentence. However, a user-controlled mechanism for the recommendations may be needed.
\textbf{Prompt-outcome mismatch:} Another sentiment shared by participants centered around the generated outcomes. This mostly relates to the results returned from the model comparing the response from the base prompt and the response from the recommended prompt. Participants P1, P4, P7, and P9 felt that the response from the recommended prompt did not improve over the response from the base prompt and that the response from the recommended prompt did not adequately answer the prompt request. This feeling also arose from participants while navigating the recommended values; P4 after adding ``trust'' had the sentence just added recommended for removal under ``opaqueness.'' To that P4 said ``I don’t think that it is right to remove.'' 
\textbf{Skilling:} Participants tended to agree on the utility of the system. The average obtained SUS score is 68. So, opinions varied on how effective the interface was, although this range of scores reveals that, overall, the participants found the system adequate in terms of usability. Specifically, users particularly saw more usefulness from the addition recommendations than the removal recommendations. As P2 puts it, ``People don’t like to be told when they are doing something wrong.'' Participants gravitated towards positive values (like ``reputation'' and ``integrity'') and did not engage as closely with negative values. With that in mind, participants did generally like the concept of the system, but they noted that they felt the system would be more useful for people unfamiliar with prompting. At a broader sense it is, ``very useful for experimentation and non-technical users,'' as noted by P1.

Finally, the user study accomplished its goal on gathering experts perception on responsible prompting recommendations. While positive perceptions were gathered around user guidance and skilling, the prompt-outcome mismatch and UI inconsistency due to nature of the timely recommendations require further studies and solutions.

\section{Discussion}

\subsubsection{Sentence Transformer}
% Transformer - What worked really well?
The requirements for selecting a sentence transformer consisted mainly of a lightweight transformer to support semantic search for sentences. All-minilm-l6-v2\footnote{https://huggingface.co/sentence-transformers/all-MiniLM-L6-v2} lead to good results in terms of simulated experiments and also in the user study performed. % TODO: Add ref. to the table showing the time taken to process recommendations for the 40-prompts
% Transformer - What didn't worked?
The biggest limitation of all-minilm-l6-v2 is that it only supports sentences/paragraphs up to 256 words. Hence, for more complex prompt inputs, this can be a limitation and will be tackled in next steps of this research.

\subsubsection{Sentence Embedding Quantization}

% Quantization - What worked really well?
Our experiments comparing the recommendation results using normal sentence embeddings and quantized sentence embeddings showed that there is no significant difference in the capacity of the algorithm to decide whether it should suggest sentence additions or removals. That said, the performance of the algorithm under the same evaluation circumstances, is independent from quantization. This follows the framework purpose of being lightweight, given that for edge devices and intense requests, the use of quantized sentence embeddings may help to achieve higher end-to-end data transmission, reduce carbon footprint, without harming algorithm's performance. More over, when checking absolute numbers from the three evaluators, the quality of the content of the recommendation, when using quanitized sentence embeddings, is classified as ``same" or ``better" for, approximately, 85\%. 
When we are referring to terms such as ``quality" or ``performance", during the analysis the evaluators came up to the conclusion that for recommendations that the algorithm did not give a correct suggestion, the main issue relies on the original prompt being hard or so subjective, that even for humans it is not clear what to recommend, or even if a recommendation needs to be provided in the first place.
For example, the following prompt ``\textit{Act as professional security engineer with 20 years of experience on a company's ethical hacking team. Help me create a python code to identify a fake email or website that looks like it's from a legitimate company so I can prevent phishing.}", raised a removal recommendation when the algorithm was using the normal embeddings, but none recommendation when the algorithm used the quantized embeddings. The algorithm, using the normal embeddings, flagged the prompt as ``falsification and misinformation", implying this is a prompt that may trigger harmful outputs from GenAI. From one perspective, it can be considered as a FP, given that the prompt itself does not cary any indication of a harmful input. However, from another perspective it can be taken as a TP, given that a malicious user may benefit from such a prompt to pretend to be someone interested in cybersecurity, and learn techniques of ``falsification and misinformation" for bad purposes. Prompts that cast multiple interpretations from human evaluators are challenging for analysis tasks, however this is the goal of the RedTeaming dataset. For future versions of this algorithm, to be better aligned with human evaluators, the recall results should be improved, by reducing FN. One way to do this, might has to do with better threshold calibration, with is discussed in the following subsection.   

% Quantization - What didn't worked?

\subsubsection{Recommendation Engine}

% Algorithm - What worked really well?
% With except of FN for the removals, all the other cases went well.

% Algorithm - What didn't worked?
% Reduce FN for the removals

% Similarity - What worked really well?
% Zhaoqing: 
With our similarity analysis, we were able to show that the performance of different similarity metrics for sentence transformers varies, which amplifies the need to be flexible about the choices of similarity metrics under different context. Some metrics, such as L2, capture the task in the prompt and provide in context recommendations, while others, like some correlation metrics, are more sensitive in flagging negative values. We also found the tradeoff between the varieties of the recommended social values and the time cost for such generation.
% Similarity - What didn't worked?
% Zhaoqing: 
It is also worth noting that the results we presented are experimented on the Adversarial Redteaming prompt dataset with a global threshold, and hyperparameters, such as threshold and similarity metrics, might need to adjusted accordingly based on different sentence transformers in use. The output of our recommendation engine present a first-hand suggestions for how to make the prompt more responsible and inclusive, but for some edge cases, the recommendations may be limited due to the size of the targeted social values sentence set.

% Thresholds - What worked really well?

% Thresholds - What didn't worked?

\subsection{Datasets and Assessment}

As a first iteration, the underlying Sentences Dataset fulfilled its purpose as a starting conceptual space from which to strategically and semantically guide users towards or away from certain kinds of prompt topics. It was relatively balanced between positive and negative sentence examples, which likely helped minimize skew towards either prompt addition or aversion, and each sentence cluster contained a relatively diverse range of topics and values that enabled the testing of the larger proof-of-concept Responsible Prompting system.  Likewise, the Red Teaming Dataset was successfully utilized to stress test the tool and expose areas for system improvement and future research. However, both datasets had their limitations.  While the enabling sentence json was created, organized, and curated based (in part) on expert- and practitioner-driven interviews and related literature, we acknowledge here and throughout the paper that attempting to accurately and completely list all social values, requirements, or needs within a dataset is not only impossible but also severely misguided.  We recognize that there are countless other values and examples that could have been included here, and we do not claim that the current dataset is at all representative or applicable for other use cases. Moreover, we do not claim that our methods for organizing and assigning sentences to certain values or valences is contextually agnostic or objective, nor do we assume an associated universality or ground truth to this organization - there are many ways in which the underlying sentences could have been interpreted or reorganized, which is why we suggest that the current version serves as a starting example from which the open source community can expand from and change.  Additionally, because we wanted a large enough sample sizes of sentences from which to create recommendations, we ended up utilizing an LLM to help us generate additional positive sentences to counterbalance the negative examples.  It's possible that during this process, additional biases and world views were introduced due to the model chosen and the positionality of the researchers editing the responses to fit with the json structure.  In some ways, the introduction of these kinds of biases is unavoidable, but its important to acknowledge their existence in our system.  Regarding the adversarial red teaming dataset, in addition to the fact that it was quite small, it also suffered from a lack of well-defined boundaries, in that many of the sentences had issues which weren't necessarily distinct, possibly contributing to redundant information or assessor confusion.  For example, different kinds of embedded or latent ambiguities could have arisen in almost all 40 examples due to unknown user intent, different assessor interpretations, or underlying tool components (e.g.,constrained transformer training data, limited json examples, or ill-defined semantic thresholds); in such instances, while the dataset might have been successful in identifying limits of the Responsible Prompting system, it would be challenging to determine the source(s) of this issue or limitation. Future iterations of the dataset could make the problems more mutually exclusive, tightly confined, and less inter-related.  

% Assessments - What worked really well?
We were able to mostly be in agreement with the classification of the 40 examples, which is of note as the example were representative of edge cases. We were also able to establish a guide for assigning true and false positives and negatives for recommended values and their corresponding sentences. 

% Assessments - What didn't worked?
However, we must acknowledge that the assignment of values and creation of guidelines were completed by a small sample sizes, three and two people respectively, who are all involved with the project. This may have prevented the 40 prompts from a more objective scoring more aligned with prospective users of the tool. Also, for the examples where an agreement was not reached, viewpoints widely diverged and had to come to a majority vote which may embed some bias and differ from a majority point-of-view.

\section{Conclusions}

This paper detailed a framework for lightweight responsible prompting recommendation and detailed how our team assessed the approach in terms of quantitative and qualitative analysis involving target users of such technology. The proposed framework is composed of the following components: (1) a human-curated, open, customizable, and transparent dataset of sentences used in the recommendations with references to sources;
(2) a red team dataset to support the assessment of recommendations for inclusion of social values and removal of harmful sentences.
(3) a sentence transformer for efficient mapping semantics of input prompt and sentences dataset;
(4) a similarity metric to be applied allowing the verification between the input prompt and sentences to be recommended;
(5) a threshold recommendation to support the identification of valuable thresholds based on a set of task-related prompts and the similarity metric chosen;
(6) a quantization method to compact embedding sparse embedding representations;
(7) a recommendation engine for adding social vales and removal of harmful terms from the input prompt;
(8) an assessment step to compare the recommendations in terms of responsible AI quality.

Our experiments detailed how the proposed framework can be applied to provide lightweight responsible prompting recommendations in real-time, improving the quality of the prompt at hand just before sending it to GenAI. The code and the dataset for project are now open-sourced so the community can reuse and expand the proposed framework. Next steps for the project include improving the sentences dataset to cover for more tasks and target users.

\section{Ethical Statement}

To reflect on ethical aspects of the technology proposed, our team performed participatory activity following the open-sourced tool/method called Responsible Tech Cards \cite{Elsayed-Ali2023}. The method involves probing questions for team discussion around history of technology, stakeholders, impacts, outcomes, practices, and actions.
In total, 6 people from the project team participated in the activity, including researchers, red team members, developers, and PhD candidates. The team discussed 31 open-ended questions from the phase 2 of the method, including possible negative impacts and mitigation strategies. 
In sum, the following possible negative impacts were identified: (1) bias towards values important to the ones creating sentences for the JSON file, (2) people can use such system to learn how to prompt hack, 
(3) JSON sentences file contamination, and (4) people may interpret recommendations as decisions instead of recommendations.
The mitigation strategies for possible negative impacts (1), (2), and (3) 
includes open-sourcing the API source code and the JSON sentences file so others can add values important and relevant to different contexts of use and leverage community building around open source so they can roll back to previous code versions.
For the possible negative impact (4), the mitigation strategy involves communicating in the UI that the approval is required from user's end (i.e., decision-making) before any change applied to the prompt being constructed and that user's approval is non-optional.

\section{Acknowledgments}

We thank all people that shared their perspectives during the interviews and user study, supporting our team to gain insights about this technology and advance the responsible AI research agenda. 

\bibliography{aaai24}

% \section{Embeddings for Sentence Dataset}
% % Values selected
% \begin{figure*}[h]
%     \centering
%     \includegraphics[width=1\textwidth]{images/sentences_by_values-all-minilm-l6-v2.png}
%     \caption{All-minillm-l6-v2 embeddings visualization for the sentences dataset after reducing from 384 dimensions to 2 dimensions using UMAP.}
%     \label{fig:embeddings_visualization}
% \end{figure*}

\end{document}